\documentstyle[11pt,newpasp,twoside,epsf,psfig]{article}
\def\edcomment#1{\iffalse\marginpar{\raggedright\sl#1\/}\else\relax\fi}
\reversemarginpar

\begin{document}

\title{Evolution of massive and magnetized protoplanetary disks}
 \author{S\'ebastien Fromang, Caroline Terquem,}
\affil{Institut d'Astrophysique de Paris, 98Bis Bd Arago,75014 Paris}
\author{Steve A. Balbus \altaffilmark{1} \&
Jean-Pierre De Villiers }
\affil{Virginia Institute of Theoretical Astronomy, Department of
Astronomy, University of Virginia, Charlottesville, VA 22903-0818}

\altaffiltext{1}{Laboratoire de Radioastronomie, \'Ecole
Normale Sup\'erieure, 24 rue Lhomond, 75231 Paris CEDEX 05, France} 

\begin{abstract}
We present global 2D and 3D simulations of self-gravitating magnetized
tori. We used the 2D calculations to demonstrate that the properties
of the MRI are not affected by the presence of self-gravity: 
MHD turbulence and 
enhanced angular momentum transport follow the linear growth of the
instability. In 3D, we have studied the interaction between an 
$m=2$ gravitational 
instability and MHD turbulence. We found its strength to be
significantly decreased by the presence of the latter, showing that both
instabilities strongly
interact in their non-linear phases. We discuss the consequences of
these
results.
\end{abstract}

\section{Introduction}

In the early phases of star formation, the forming accretion disks are
likely to be very massive because of a strong infall from the parent
molecular cloud. These massive disks are subject to the development of
gravitational instabilities which transport angular momentum outward
(Laughlin, Korchagin, \& Adams 1998). In addition, when sufficiently
ionized, the disks are also
unstable to the MRI (Balbus \& Hawley 1991, 1998). The simultaneous
development
of these instabilities in disks may significantly affect their 
evolution. We have undertaken a study of self-gravitating
magnetized tori
by means of numerical simulations.

In section 2, we describe the numerical methods we have used. 
In section 3, we present the
results of the 2D simulations, focusing on the properties of the MRI in
a self-gravitating environment. In section 4, we review preliminary
results obtained in 3D calculations, and we discuss the implications of
our results in section 5.

\section{Numerical methods}

We used the code Zeus-2D (Stone \& Norman 1992a, 1992b) to perform the
2D calculations and the code GLOBAL (Hawley \& Stone 1995) for the 3D
simulations. Both codes solve the MHD equations using time-explicit
Eulerian finite differencing and the Constrained Transport method to
evolve the magnetic field.  The Poisson solver used to calculate the
gravitational potential $\Phi_g$ involves two steps: we first
calculate $\Phi_g$ on the boundary, using the Legendre functions
well-suited to the cylindrical geometry (Cohl 1999), and we apply the
Successive Over-Relaxation method to update $\Phi_g$ everywhere on the
grid (Hirsch 1988).

\section{2D simulations}

We have built an initial equilibrium structure using the
Self--Consistent Field (SCF) method (Hachisu 1986). The inner and
outer radii of the torus are $R_{in}=0.3$ and $R_{out}=1$
respectively. We set the angular velocity $\Omega \propto r^{-1.68}$,
where $r$ is the cylindrical radius.  A central mass having half the
mass of the disk is present. The equation of state is adiabatic.

In the following, we compare the evolution of this torus to its zero
mass counterpart, the gravitational potential of the latter being only
that of a central point mass. At the beginning of the simulation, a weak
poloidal magnetic field is added to the equilibrium structure, with the
toroidal component of the vector potential being:

\begin{equation}
A_{\phi} \propto \rho \cos \left( 2\pi
\frac{r-R_{in}}{R_{out}-R_{in}} \right)
\label{Aphi}
\end{equation}

\noindent where $\rho$ is the mass density.  The components of the
magnetic field are then scaled such that the volume averaged ratio of
gas to magnetic pressure (hereafter called $\langle \beta\rangle$)
equals $1500$ for the self-gravitating model and $200$ for the zero
mass one. The resolution in both model is $(N_r,N_z)=(256,256)$ in the
radial and vertical directions, respectively.

We found the MRI grows in both models, developing approximately the
same Maxwell stress (defined by $T^{Max}_{r\phi}=-B_r B_{\phi}/4\pi$).
In the self-gravitating case, the evolution is very similar to what
was found in previous non self--gravitating calculations (e.g. Hawley
[2000)].  Namely, the initial linear growth of the instability is
followed by a turbulent phase during which angular momentum is
transported outward. Turbulence then gradually decays because of the
anti-dynamo theorem.

\begin{figure}
\plottwo{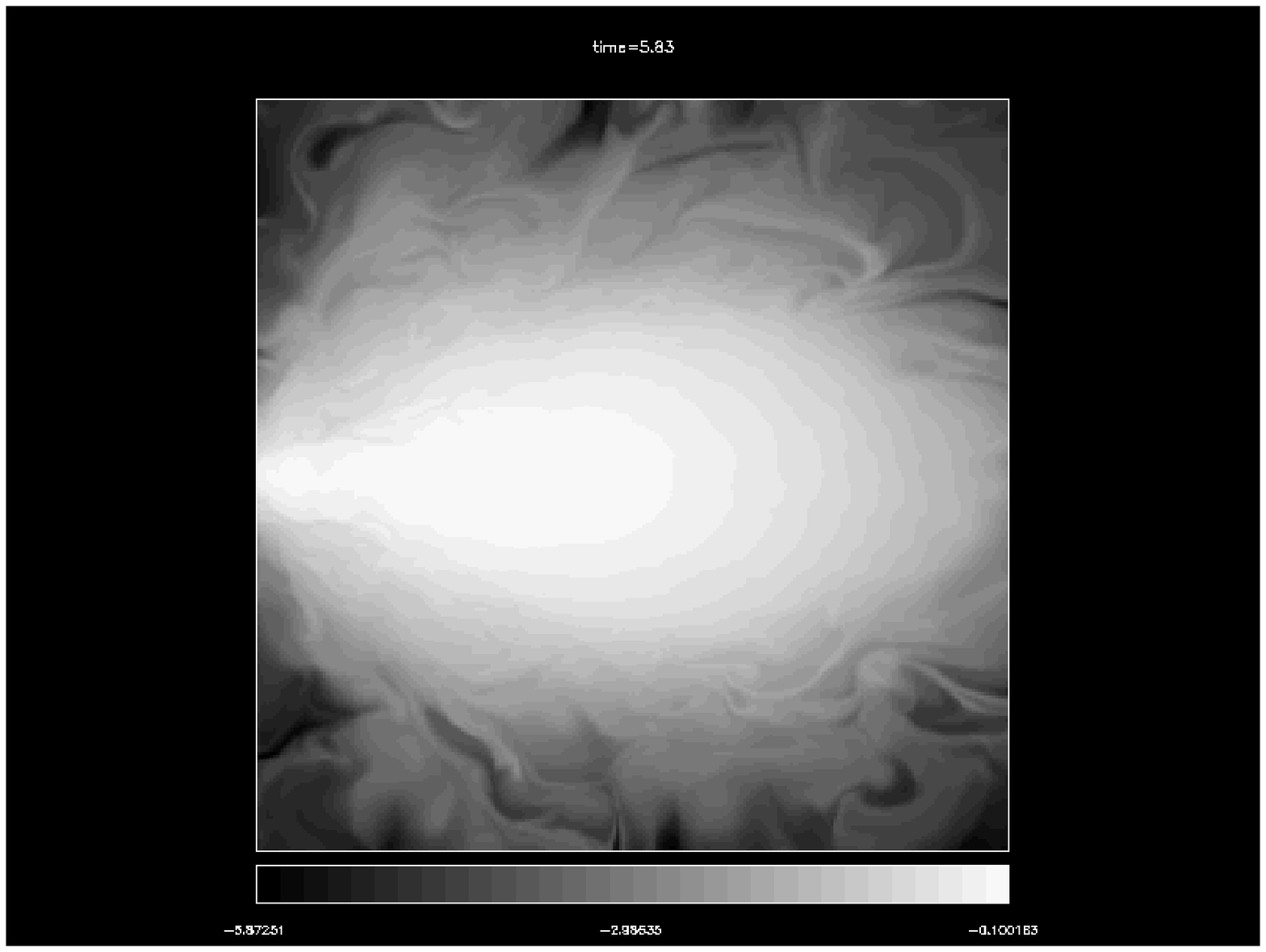}{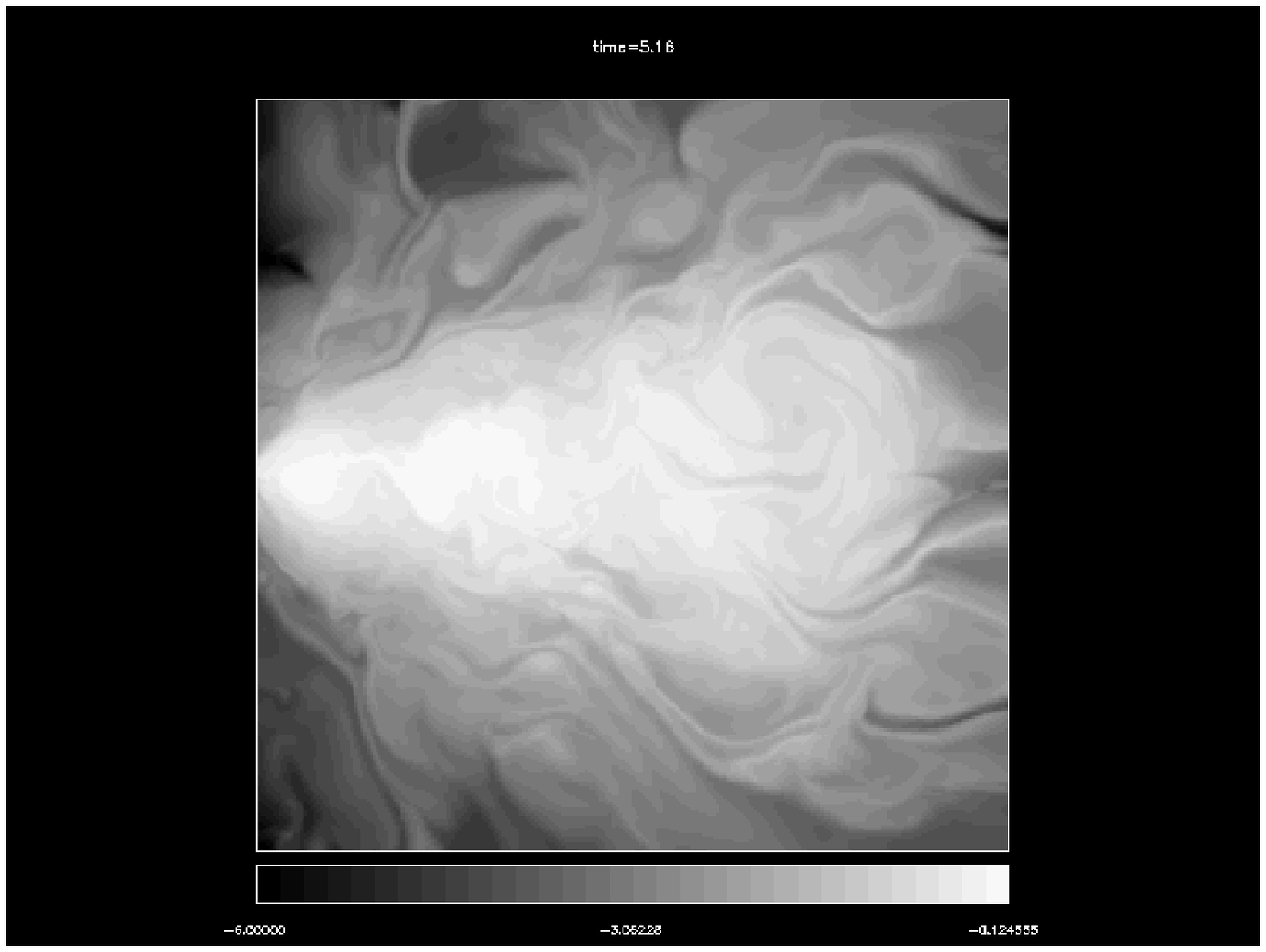}
\caption{Comparison between the logarithm of the density in the
($r$-$z$) plane for the self-gravitating torus ({\it left panel}) and
the zero mass torus ({\it right panel}). The former develops a
two-component structure composed of an inner Keplerian disks and an
outer, more massive, thick disk. The later is disrupted by the MRI, and
a standard thin disk structure builds up, with a constant $H/R$ ratio.}
\label{snapshots}
\end{figure}

\begin{figure}[t]
\plotfiddle{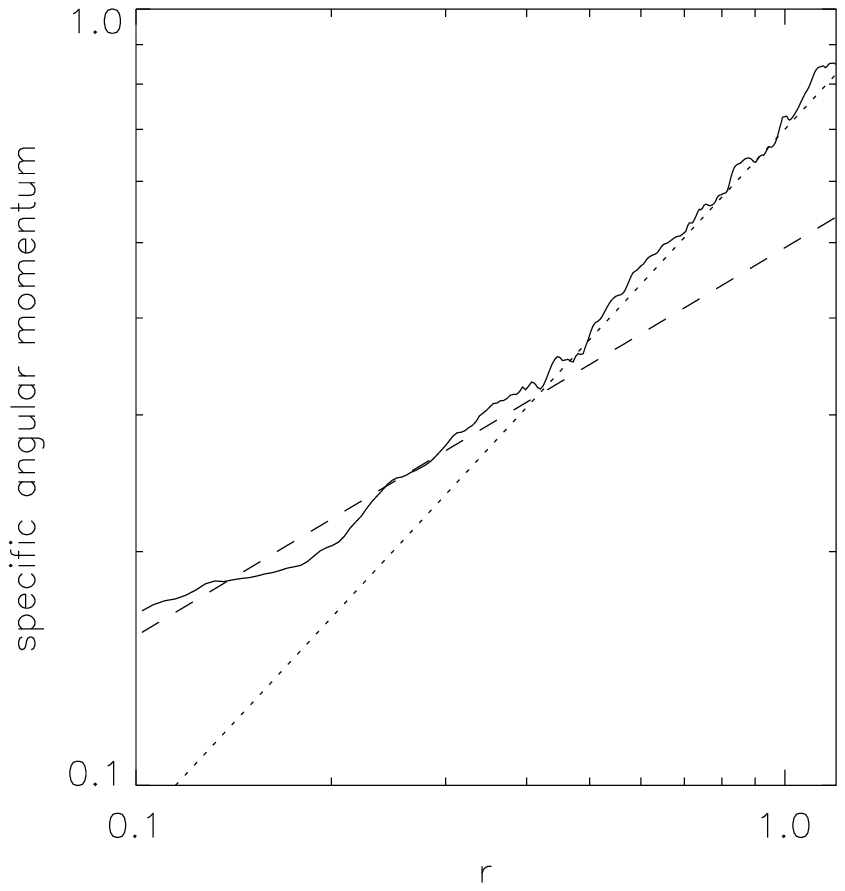}{60mm} {0} {70} {70} {-150} {-270}
\caption{Angular momentum profile in the equatorial plane of the
torus,
averaged between orbits 5.05 and 7. The dashed line shows the Keplerian
profile
resulting from the central point mass. The dotted line has a power law
dependence with radius: $l \propto r^{0.9}$.}
\label{ang_mom}
\end{figure}

We show in figure $1$ the logarithm of the density field in the
$(r-z)$ plane during the turbulent phase for both models. In the
self-gravitating case ({\it left panel}), the initial torus has
developed a two-component structure, composed of an inner thin disk
fed by an outer thick, massive torus. Figure $2$ shows the angular
momentum $l$ radial profile in the equatorial plane during this phase
({\it solid line}). As shown by the dashed line, the inner disk is in
Keplerian rotation around the central mass.  The dotted line is a fit
of the outer part of the disk with a power law dependence $l \propto
r^{0.9}$, very close to the Mestel profile $l \propto r$.
   
Coming back to figure $1$, the comparison between both models is
striking: although the Maxwell stress is similar in the two cases, the
self-gravitating torus presents a much more featureless structure than
its zero mass counterpart.  This result is due to the large scale
coherence of the self-gravitating potential, which smoothes MRI
fluctuations.  We shall see that this effect persists in 3D.

\section{3D simulations}

\begin{figure}
\plotfiddle{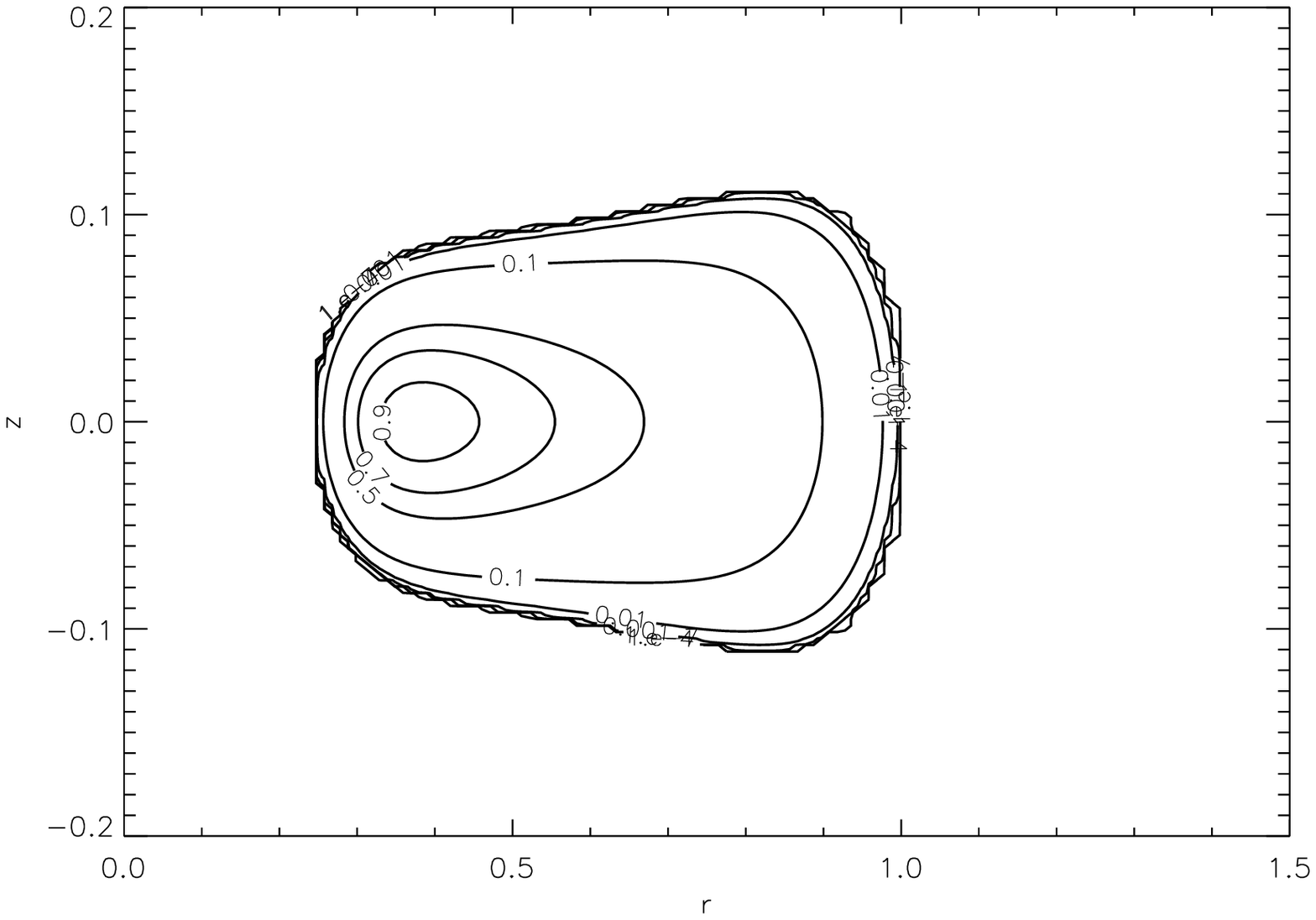}{20mm} {0} {40} {40} {-190} {-215}
\plotfiddle{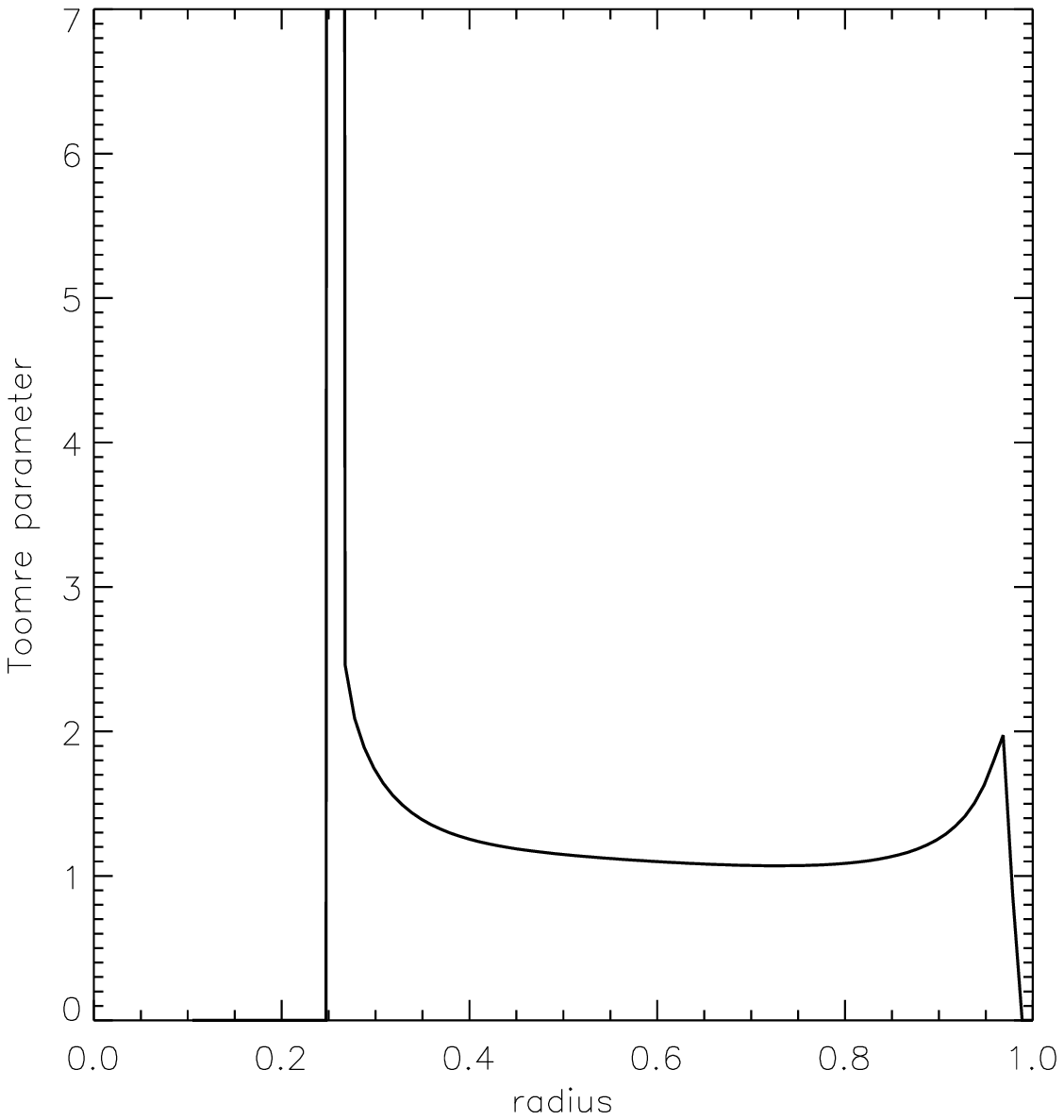}{20mm} {0} {40} {40} {30} {-5}
\caption{Density contours for the disk model used in the 3D simulations
(left panel). The contours shown are
$\rho=10^{-7},10^{-4},0.001,0.01,0.1,0.5,0.7$ and $0.9$. The right panel
shows the radial profile of the Toomre $Q$ parameter.}
\end{figure}

In this section, we present the results of our 3D simulations.  As
above, we have used the SCF method to compute the disk model, but with
different parameters.  Here the central mass is twice that of the
disk, and the specific angular momentum $l$ in the disk is chosen such
that

\begin{equation}
l \propto m^2(r) \, ,
\end{equation}

\noindent
where $m(r)$ is the mass inside the radius $r$ divided by the total disk
mass.  The density contours of this disk model are shown in figure $3$
(left panel).  We introduce the Toomre parameter $Q$ defined by:

\begin{equation}
Q=\frac{c_s\kappa}{\pi G \Sigma} \, ,
\end{equation}

\noindent
where $c_s$ is the sound speed, $\kappa$ the epicyclic frequency and
$\Sigma$ the surface density of the disk. Linear studies have shown
that a disk becomes gravitationally unstable when $Q$ is of the order
of $1$.  The radial profile of $Q$ in our disk model is shown on the
right panel of figure $3$.  We have $Q \sim 1$ over a large range of
radii.  We therefore expect the disk to be gravitationally unstable.

\subsection{Hydrodynamical simulations}

\begin{figure}[t]
\plotfiddle{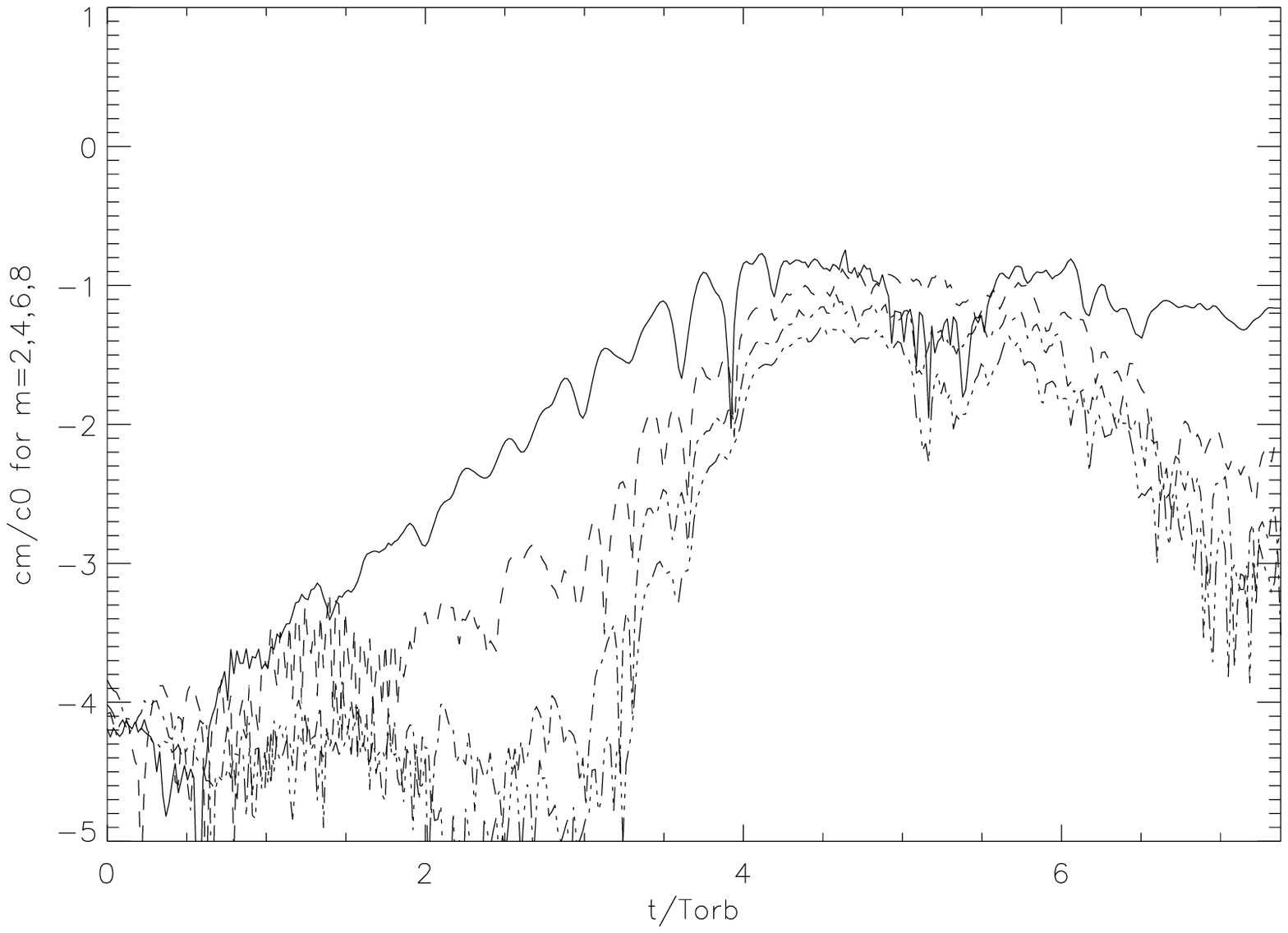}{40mm} {0} {40} {40} {-130} {-160}
\caption{Time evolution of the Fourier components of the density in the
equatorial plane for the modes $m=2,4,6,8$ (from top to bottom). The
$m=2$ mode is unstable.}
\end{figure}

\begin{figure}[t]
\plotfiddle{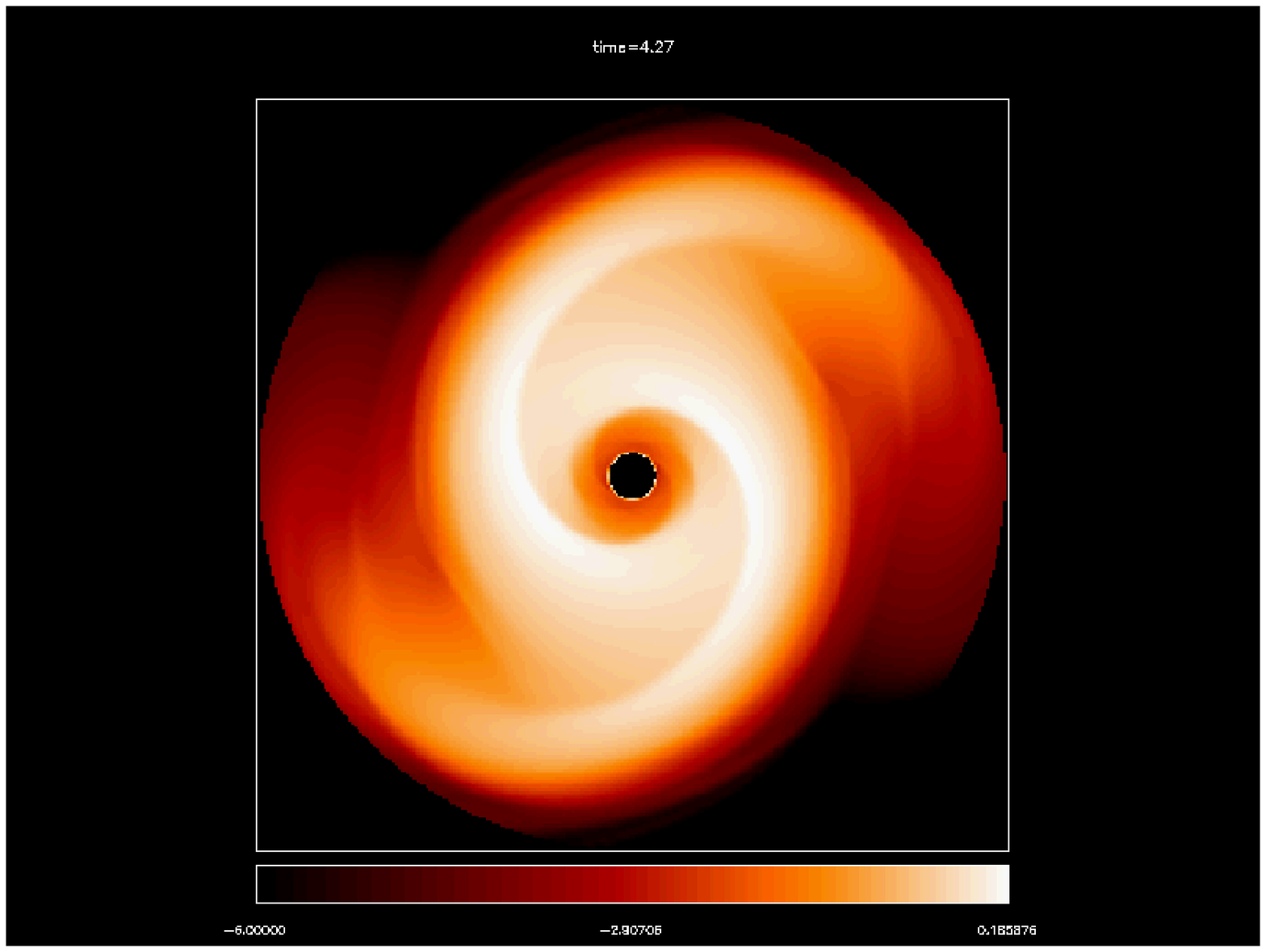}{50mm} {0} {40} {40} {-120} {-150}
\caption{Logarithm of the density in the equatorial plane after $4.27$
orbits.}
\end{figure}

We have performed hydrodynamical simulations to check that our code
behaves as expected in this regime.  In the simulations presented here,
the resolution is $(N_r,N_{\phi},N_z)=(128,64,64)$ in cylindrical
coordinates. The azimuthal domain is restricted to $\phi \in [0,\pi]$,
which prevents the growth of modes with odd values of the azimuthal
number $m$.
Time is measured in units of the orbital period at the initial outer
edge of the disk.

The time evolution of the Fourier component of the density in the
equatorial plane is shown in figure $4$ for the modes $m=2,4,6,8$
(from top to bottom). The $m=2$ mode grows at the beginning of the
simulation and saturates after $4$ orbits. The development of this 
2~arms spiral structure can be seen on figure~$5$, which shows the
logarithm of the density in the equatorial plane after $4.27$ orbits.

During the simulation, matter is driven toward the center of the disk
by the gravitational torque associated with the spiral arms.  As the
surface density in the disk decreases, the Toomre $Q$ parameter
increases.  As a result, the instability dies away after 8~orbits.

These results are in agreement with earlier simulations of similar
disks.

\subsection{3D MHD simulations in an axisymmetric potential}

\begin{figure}
\plotone{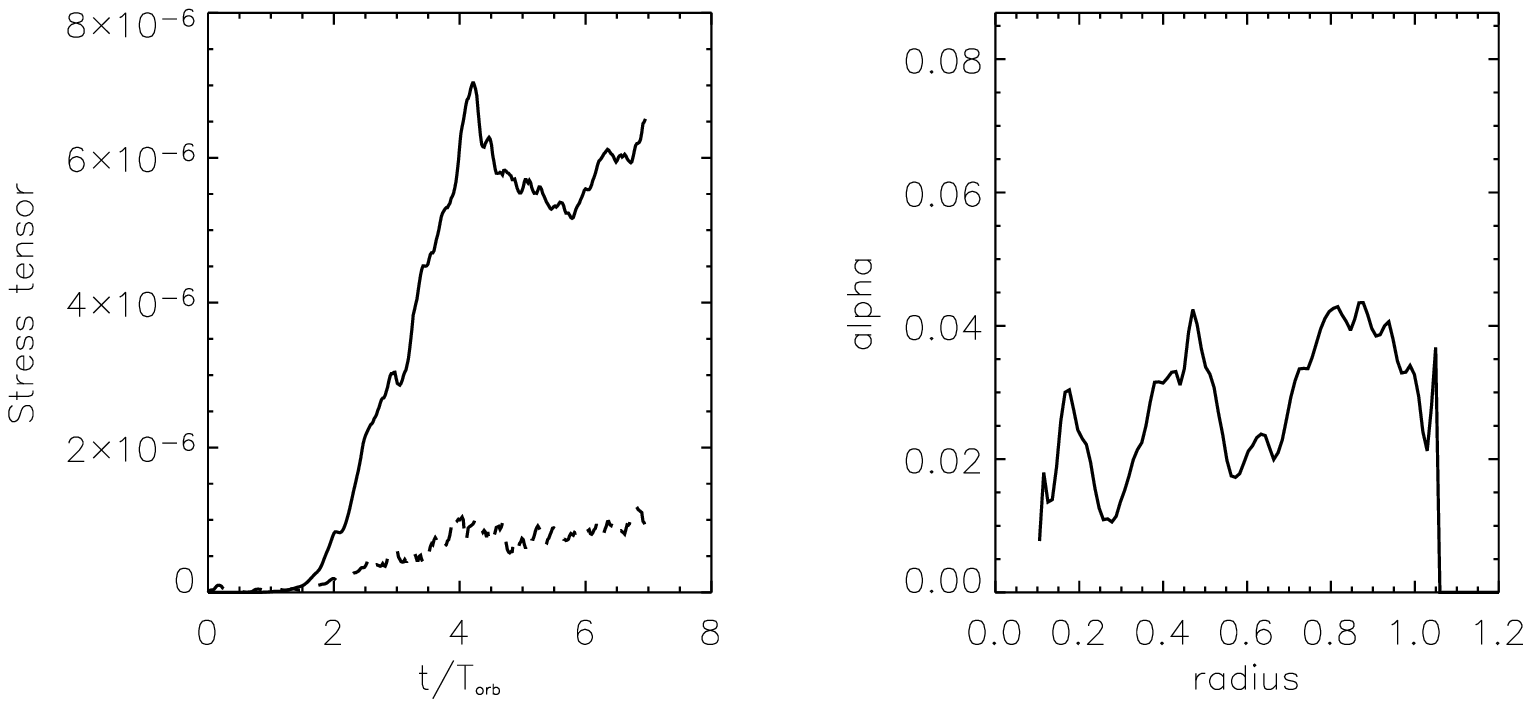}
\caption{Left panel: time evolution of the Maxwell ({\it solid line})
and Reynolds ({\it dashed line}) stress tensor in the MHD run with an
axisymmetric gravitational potential. Right panel: radial profile of
the  $\alpha$ parameter.}
\end{figure}


Next, we have performed a 3D MHD calculation of the evolution of the
disk in an axisymmetric gravitational potential, i.e. a potential for
which we have retained only the $m=0$ component.  This prevents the
growth of gravitational instabilities and enables us to check the
behavior of our code in the MHD regime.  The resolution used for this
run is $(N_r,N_{\phi},N_z)=(128,32,128)$ and the azimuthal domain is
$[0,\pi/2]$.  A toroidal magnetic field with $\langle \beta \rangle=8$
is added to the equilibrium model.  

The left panel of figure~6 shows the time evolution of the volume
averaged Maxwell and Reynolds stress tensors.  The Maxwell stress
rises during the early part of the evolution (corresponding to the
linear development of the instability) before the MRI breaks down into
turbulence after 4~orbits.  The Reynolds stress is smaller during the
whole simulation.  The right panel of figure~6 shows the radial
profile of the $\alpha$ parameter (defined as the ratio of the total
stress to the thermal pressure) at the end of the simulation.  It has
a typical value of a few times $10^{-2}$, similar to values seen in 3D
simulations of zero mass disks starting with similar magnetic field
configurations (Steinacker \& Papaloizou 2002).  

\subsection{Full 3D simulations}

\begin{figure}[t]
\plottwo{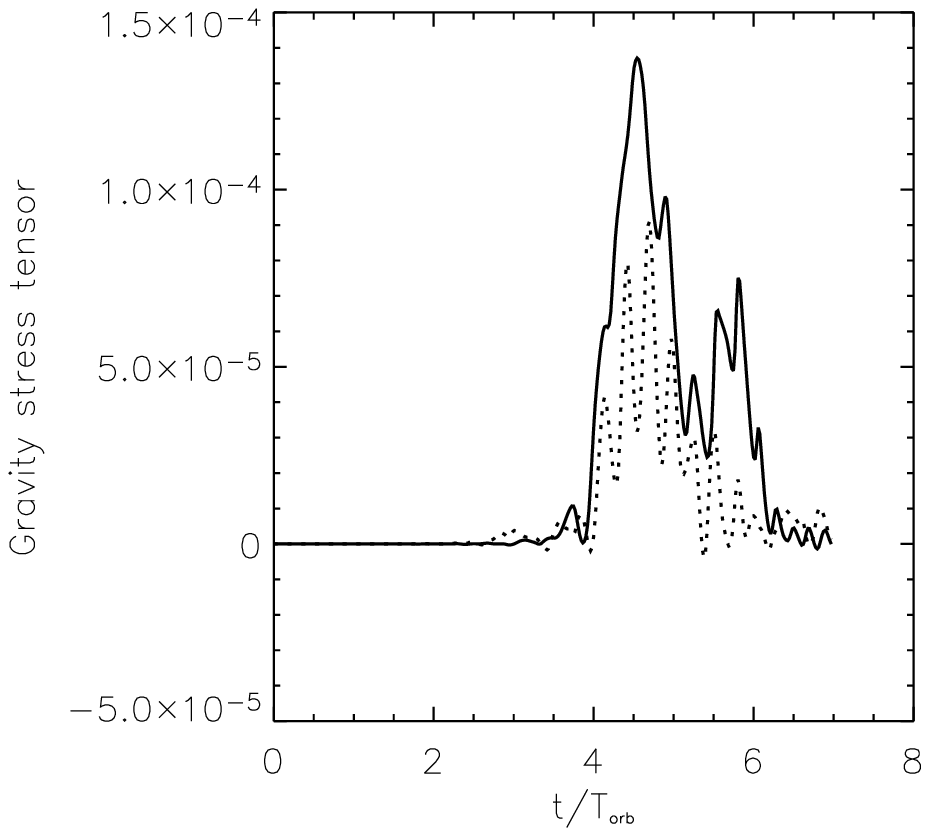}{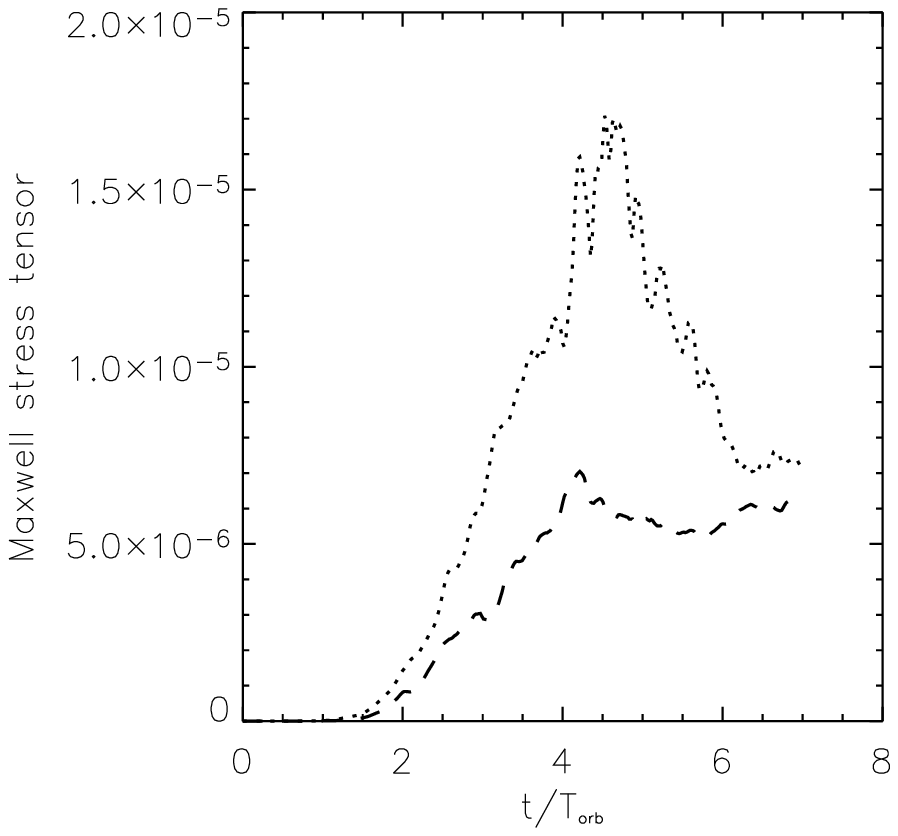}
\caption{Left panel: time evolution of the gravitational stress tensor
in the hydrodynamical run ({\it solid line}) and in the full simulation
({\it dotted line}). Right panel: time evolution of the Maxwell stress
tensor in the full simulation ({\it dotted line}) and in the 3D MHD run
with an axisymmetric gravitational potential ({\it dashed line}).}
\end{figure}


In this section, we present the results of the simulation in which
both the MRI and gravitational instabilities can develop.  We include
the Fourier components of the gravitational potential up to $m=8$.
The toroidal magnetic field that we add to the equilibrium structure
is the same as that described in the previous section.  For this
calculation, the resolution is $(N_r,N_{\phi},N_z)=(128,64,128)$ and
the azimuthal domain is $[0,\pi]$.

As expected, we observe the simultaneous appearance of MHD turbulence
along with the $m=2$ spiral arm reminiscent of the hydrodynamical
calculation. To analyze the angular momentum transport properties, we
compare the time evolution of the volume averaged gravitational and
Maxwell stress tensors obtained in this simulation with that obtained
in the runs described above.  Note that the gravitational stress
tensor is defined by:

\begin{equation}
T_{r\phi}^{grav}=\frac{1}{4\pi G}\int \frac{1}{r}\frac{\partial
\Phi_g}{\partial r}\frac{\partial \Phi_g}{\partial \phi}d\tau \, .
\end{equation}

\noindent
The left panel of figure~7 shows $T_{r\phi}^{grav}$ as a function of
time for this simulation ({\it dotted line}) and during the
hydrodynamical run ({\it solid line}).  We can see that the presence
of MHD turbulence decreases the strength of the gravitational stress
by a factor of about $2$.  This result shows that MHD turbulence and
the gravitational instability strongly interact with each other.  The
time evolution of the Maxwell stress is shown on the right panel of
figure~7 for the full simulation ({\it dotted line}) and for the MHD
run in an axisymmetric potential ({\it dashed line}). We see that the
Maxwell stress increases when the gravitational instability develops.
This appears to be due to the accumulation of the magnetic field along
the spiral arms.  When the gravitational instability disappears after
about 7~orbits (because of a decrease of the mass density), the Maxwell
stress goes back to its ``turbulent'' value.

To test the sensitivity of our results to the initial field
topology, we have run the same calculation with an initial poloidal
magnetic field.  We found the same qualitative results.

\section{Conclusions and Perspectives}

We have presented here the first global simulations of massive,
magnetized disks.

Using 2D axisymmetric numerical simulations, we have shown
that the properties of the MRI are similar in self--gravitating
and zero mass disks.  We observe that
these disks quickly develop a dual structure composed of an inner thin
disk in Keplerian rotation around the central mass, and a thicker
outer torus whose rotation profile is close to a Mestel profile.

We have then used 3D simulations to study the angular momentum
transport properties in disks when both MHD turbulence and
gravitational instabilities are present.  We have found that the
gravitational instability is affected by the presence of the
turbulence: the gravitational stress tensor is decreased by roughly a
factor of $2$ when compared with hydrodynamical simulations.  This
results in a smaller mass accretion rate toward the central object.
Self--gravitating disks may therefore have a longer lifetime than
previously thought.

Note that the simulations presented here use an adiabatic equation of
state. In this case, all the energy generated in shocks and
compression is locally converted into heat.  This prevents the
formation of bound objects by gravitational collapse.  The opposite
case would correspond to the use of a locally isothermal equation of
state, for which all the energy generated is immediately radiated
away.  Several authors (Mayer et al. 2002, Rice et al. 2003, Boss 1997
\& 1998) have indeed reported gravitational collapse in isothermal
calculations of disks, although this issue is still under debate
(Pickett et al. 2003).  We believe that the presence of MHD turbulence
affects the thermal balance in the disk and therefore needs to be
included.  We are currently performing calculations of isothermal
massive and magnetized disks.


\begin{thebibliography}{}

\bibitem[\protect\citeauthoryear{Balbus \& Hawley}{1991}]{balbus91}
Balbus,S., Hawley,J., 1991, {\it ApJ}, 376, 214
\bibitem[\protect\citeauthoryear{Balbus \& Hawley}{1998}]{balbus98}
Balbus,S., Hawley,J., 1998, {\it Rev. Mod Phys.}, 70, 1
\bibitem[\protect\citeauthoryear{Boss}{1997}]{boss97}
Boss,A.P., 1997, {\it Science}, 276, 1836
\bibitem[\protect\citeauthoryear{Boss}{1998}]{boss98}
Boss,A.P., 1998, {\it ApJ}, 503, 923
\bibitem[\protect\citeauthoryear{Cohl \& Tohline}{1999}]{Cohl99}
Cohl,H.,
Tohline,J., 1999, {\it ApJ}, 527, 86
\bibitem[\protect\citeauthoryear{Hawley \& Stone}{1995}]{hawley95}
Hawley,J., Stone,J., 1995, {\it Comput. Phys. Commun.}, 89, 127
\bibitem[\protect\citeauthoryear{Hawley}{2000}]{hawley00} Hawley,J.,
2000,
{\it ApJ}, 528, 462
\bibitem[\protect\citeauthoryear{Hachisu}{1986}]{hachisu86} Hachisu,I.,
1986, {\it ApJS}, 62, 461
\bibitem[\protect\citeauthoryear{Hirsch}{1988}]{hirsch88} Hirsch,C.,
{\it
Numerical Computation of Internal and External Flows - Volume 1,
Fundamentals of Numerical Discretization.} Wiley (1988)
\bibitem[\protect\citeauthoryear{Laughlin et al.}{1998}]{laughlin98}
Laughlin.G, Korchagin,V., Adams,F.C., 1998, {\it ApJ}, 504, 945
\bibitem[\protect\citeauthoryear{Mayer et al.}{2002}]{mayer02}
Mayer,L., Quinn,T., Wadsley,J., Stadel,J., 2002, {\it Science}, 298
\bibitem[\protect\citeauthoryear{Stone \& Norman}{1992a,b}]{Stone92a}
Stone,J., Norman,M., 1992a, {\it ApJS}, 80, 753
\bibitem[\protect\citeauthoryear{Stone \& Norman}{1992b}]{Stone92b}
Stone,J., Norman,M., 1992b, {\it ApJS}, 80, 791
\bibitem[\protect\citeauthoryear{Steinacker \&
Papaloizou}{2002}]{steinacker02} Steinacker,A., Papaloizou,J., 2002,
{\it
ApJ}, 571, 413
Press,W.H.,
\bibitem[\protect\citeauthoryear{Rice et al.}{2003}]{steinacker02} Rice,
W. K. M., Armitage, P. J., Bate, M. R., Bonnell, I. A., 2003, MNRAS,
339, 1025
\end{thebibliography}
\end{document}